\documentclass[traditabstract]{aa}
\usepackage{natbib,twoopt}
 \bibpunct{(}{)}{;}{a}{}{,}    
\usepackage{graphicx}
\usepackage{ulem}
\usepackage{times}
\begin{document}

\title{Mg {\sc ii} $\lambda$2797, $\lambda$2803 emission in a large sample
of low-metallicity star-forming galaxies from SDSS DR14}
\author{N. G. \ Guseva \inst{1,2}
\and Y. I. \ Izotov \inst{1,2}
\and K. J. \ Fricke \inst{1,3}
\and C. \ Henkel \inst{1,4}
}
\offprints{N. G. Guseva, nguseva@bitp.kiev.ua}
\institute{          Max-Planck-Institut f\"ur Radioastronomie, Auf dem H\"ugel 
                     69, 53121 Bonn, Germany
\and
                     Bogolyubov Institute for Theoretical Physics, National
                     Academy of Sciences of Ukraine, 14-b Metrolohichna str.,
                     Kyiv, 03143, Ukraine
\and
                     Institut f\"ur Astrophysik, G\"ottingen Universit\"at, 
                     Friedrich-Hund-Platz 1, 37077 G\"ottingen, Germany 
\and
                     Astron. Dept., King Abdulaziz University,
                     P.O. Box 80203, Jeddah 21589, Saudi Arabia                 
}
\date{Received \hskip 2cm; Accepted}

\abstract{A large sample of Mg~{\sc ii} emitting star-forming galaxies with low
  metallicity [O/H] = log(O/H) -- log(O/H)$_{\odot}$ between --0.2 and --1.2 dex
  is constructed from  Data Release 14 of the Sloan Digital Sky
  Survey. We selected 4189 galaxies with
  Mg~{\sc ii} $\lambda$2797,  $\lambda$2803 emission lines in the redshift range
  $z$ $\sim$ 0.3 -- 1.0
  or 35\% of the total Sloan Digital Sky Survey star-forming sample
  with redshift $z$ $\ge$ 0.3.
  We study the dependence of the magnesium-to-oxygen and
 magnesium-to-neon abundance ratios on metallicity. Extrapolating
  this dependence to [Mg/Ne] = 0
  and to solar metallicity we derive a magnesium depletion of
  [Mg/Ne] $\simeq$ --0.4 (at solar metallicity).
  We prefer neon instead of oxygen to evaluate the magnesium depletion in the
  interstellar medium because neon is a noble gas and is
  not incorporated into dust, contrary to oxygen.
  Thus, we find that more massive and more metal abundant
  galaxies have higher magnesium depletion.
  The global parameters of our sample, such as the mass of the stellar
  population
  and star formation rate, are compared with
  previously obtained results from the literature.
These results confirm that Mg {\sc ii} emission has a
 nebular origin.
Our data for interstellar magnesium-to-oxygen abundance ratios
   relative to the solar value are in good agreement with
similar measurements made for Galactic stars,
for giant stars in the Milky Way satellite dwarf galaxies, and with
low-metallicity damped Lyman-alpha systems.
}
\titlerunning{The Mg {\sc ii} $\lambda$2797, $\lambda$2803 emission
in star-forming galaxies from the SDSS DR14}
\authorrunning{N. G. Guseva et al.}
\maketitle

%%%%%%%%%%%%%%%%%%%%%%%%%%%%%%%%%%%%%%%%%%%%%%%%%%%%%%%%%%%
\begin{figure*}
\hbox{
 \hspace*{0.0cm}\includegraphics[angle=-90,width=0.48\linewidth]{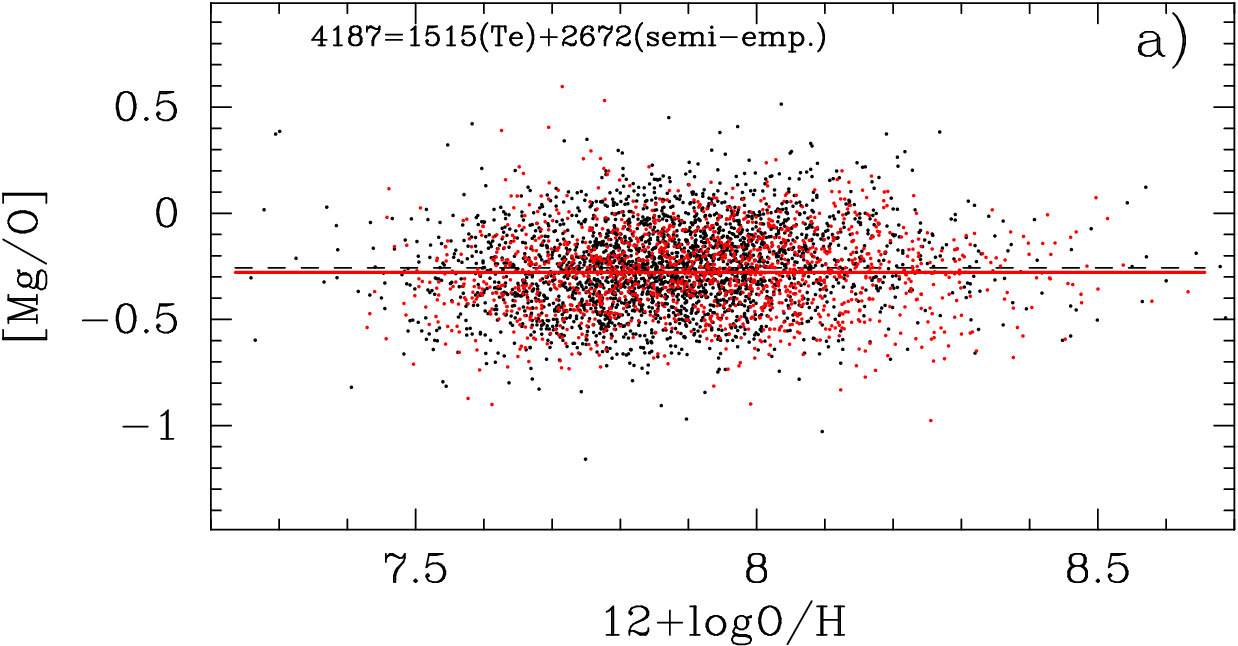}
 \hspace*{0.4cm}\includegraphics[angle=-90,width=0.48\linewidth]{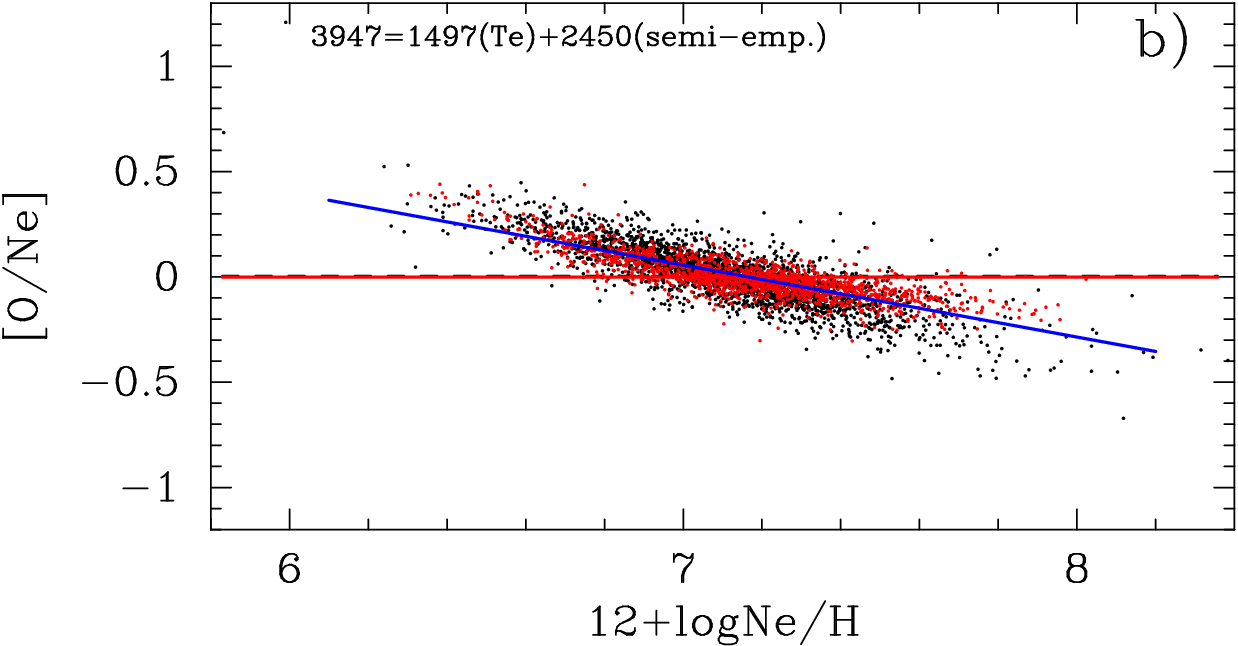}
}
\hbox{
 \hspace*{0.0cm}\includegraphics[angle=-90,width=0.48\linewidth]{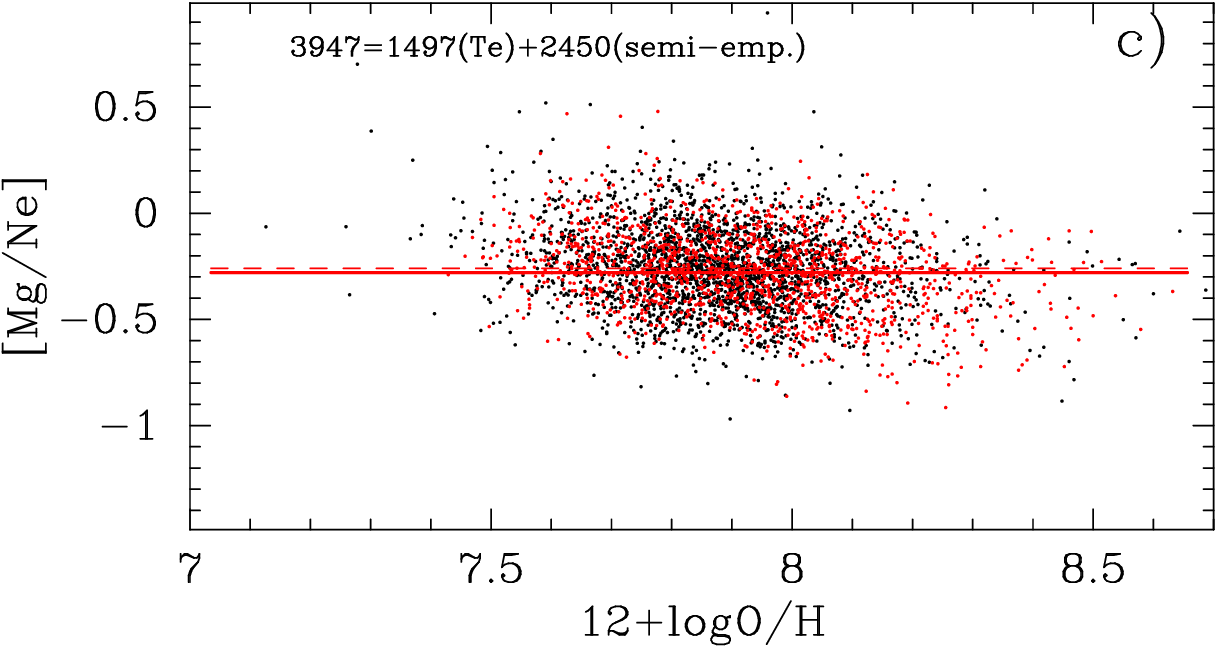}
 \hspace*{0.4cm}\includegraphics[angle=-90,width=0.48\linewidth]{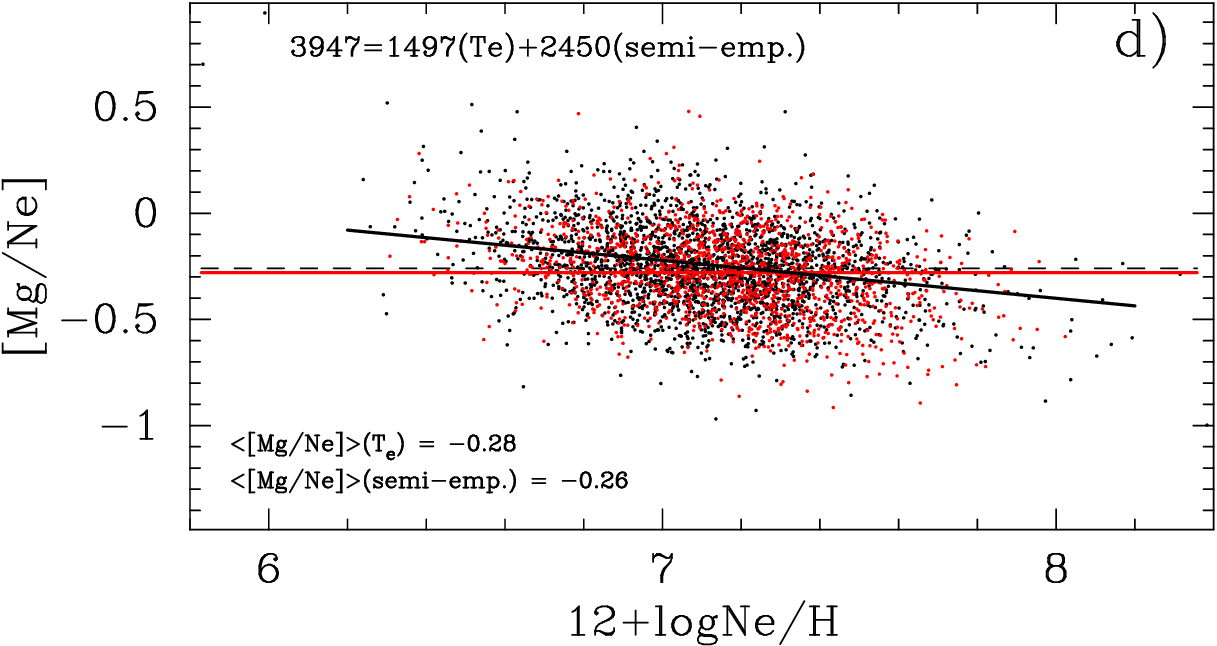}
}
\caption{\textbf{a}) Magnesium-to-oxygen abundance ratio relative to
  solar value [Mg/O] = log(Mg/O) -– log(Mg/O)$_\odot$ vs. metallicity expressed
  in 12 + log(O/H). Galaxies with metallicities derived using the direct
  $T_{\rm e}$ method are in red; those obtained by the semi-empirical method are
  in black (see Sect.~\ref{S3}).
    The average values of [Mg/O]  from the direct
  and semi-empirical methods are shown by solid
  red and dashed black lines, respectively, and are very similar (--0.279
  and --0.256). \textbf{b}) Oxygen-to-neon ratio relative to the solar value
  [O/Ne] vs. 12 + log(Ne/H). \textbf{c}) and \textbf{d}) [Mg/Ne] vs.
   12 + log(O/H) (\textbf{c}) or 12 + log(Ne/H)
   (\textbf{d}). Linear regressions are shown in \textbf{b}) and \textbf{d})
   by solid blue and black lines, respectively.
}
\label{fig1}
\end{figure*}

%%%%%%%%%%%%%%%%%%%%%%%%%%%%%%%%%%%%%%%%%%%%%%%%%%%%%%%%%%%%%

\section {Introduction}\label{S1}

Magnesium is one of the most abundant elements of stellar nucleosynthesis
produced in massive stars and the ninth most abundant element in the
Universe; it is widely
used in studies of damped Lyman-$\alpha$ absorber (DLA) systems, the local interstellar medium (LISM), the
Galactic halo and disk stars, stars in the Milky Way satellite dwarf galaxies,
and in star-forming galaxies.
It is a moderately refractory element with a condensation temperature
of $\sim$1340K \citep{SavageSembach1996}.
  Magnesium is thus an
important constituent of dust grains.
   There are several motivations for this study of Mg~{\sc ii} in
star-forming galaxies.
  Although  many studies of the Mg~{\sc ii} resonant doublet
in  emission and/or absorption have been carried out in recent years
\citep{Chen2000,Bonifacio2004,Monaco2005,Sbordone2007,Lai2008,Bensby2014,Roederer2014,Cooke2015,Cia2016,Finley2017,Henry2018,Feltre2018,Hill2018},
physical mechanisms explaining the collected data are not completely clear,
specifically for galaxies.
   Among others factors, the resonant scattering of nebular continuum photons in outflows
or resonant scattering in low ionisation or neutral gas have to be considered.
    An accurate Mg abundance determination in the interstellar medium and its
comparison to stellar magnesium abundances and to model yield predictions
allow us to estimate the quantity of magnesium as a refractory element
incorporated into interstellar dust grains.

Large surveys of galaxies such as the Sloan Digital Sky Survey (SDSS) offer a good opportunity to
substantially expand the statistics of previous studies. It is also interesting 
to extend the redshift range, the range of stellar population masses,
metallicities, star formation rates (SFRs), specific star formation rates
(sSFRs),
and types of galaxies with Mg {\sc ii} in emission and/or absorption.
A statistical increase of databases of star-forming galaxies with Mg~{\sc ii}
emission
is necessary, similar to investigations performed recently for Galactic stars
where sample sizes have grown from hundreds of stars to several hundred thousand
stars with magnesium detections \citep[e.g. the second data release of the
GALAH survey
containing more than 300 thousand stars, ][]{Buder2018,Buder2018a}.

   This study is a continuation of our earlier study of 45 Mg~{\sc ii} emission
low-metallicity star-forming galaxies with
Mg~{\sc ii} $\lambda$2797, $\lambda$2803 emission
performed on the base of SDSS/DR7 \citep{Guseva2013}.
   Now we consider many more Mg~{\sc ii} emitting galaxies selected from the
SDSS/Baryon Oscillation Spectroscopic Survey (BOSS) DR14 containing significantly more distant galaxies.

%%%%%%%%%%%%%%%%%%%%%%%%%%%%%%%%%%%%%%%%%%%%%%%%%%%%%%%%%%%%%

\begin{figure}
\hbox{
 \hspace*{0.0cm}\includegraphics[angle=-90,width=0.85\linewidth]{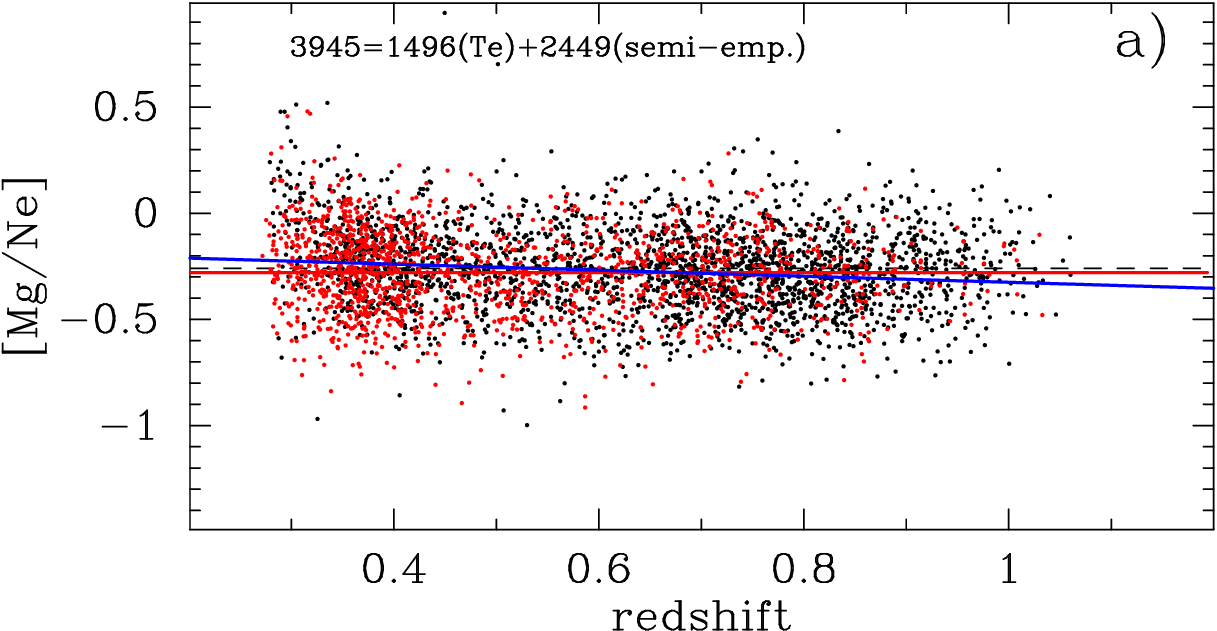}
}
\hbox{
 \hspace*{0.0cm}\includegraphics[angle=-90,width=0.85\linewidth]{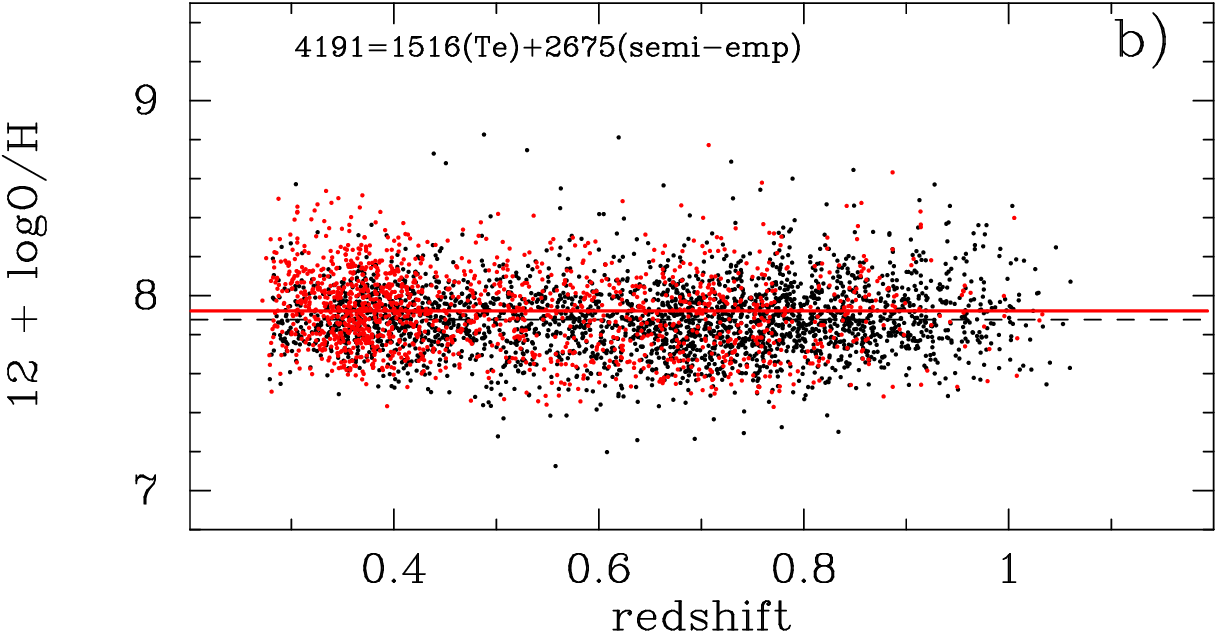}
}
\caption{\textbf{a}) Dependence of the magnesium-to-neon abundance ratio
  relative
to the solar value [Mg/Ne]  and \textbf{b}) metallicity expressed in 12 + log(O/H)
on redshift.
}
\label{fig2}
\end{figure}

%%%%%%%%%%%%%%%%%%%%%%%%%%%%%%%%%%%%%%%%%%%%%%%%%%%%%%%%%%%%%

\begin{figure}
\hbox{
 \hspace*{0.0cm}\includegraphics[angle=-90,width=0.95\linewidth]{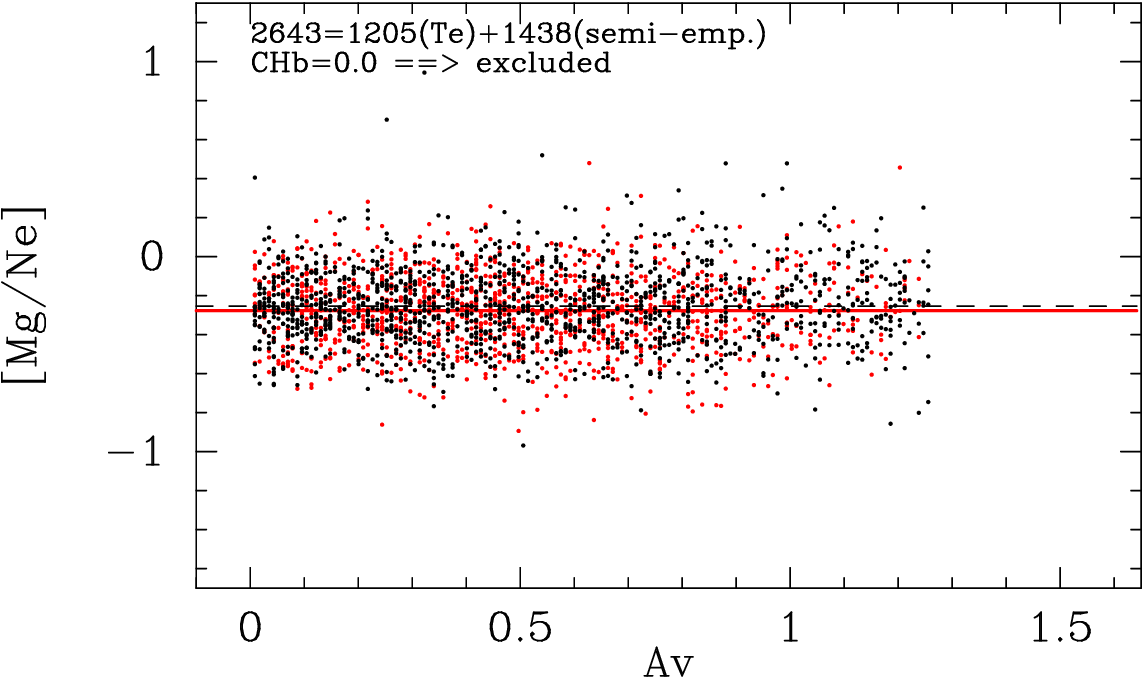}
}
\caption{Dependence of the magnesium-to-neon abundance ratio relative
to the solar value, [Mg/Ne], on extinction A$_V$ in magnitudes.
}
\label{fig3}
\end{figure}

%%%%%%%%%%%%%%%%%%%%%%%%%%%%%%%%%%%%%%%%%%%%%%%%%%%%%%%%%%%%%

\section {Sample selection \label{S2}}

  A large sample of $\sim$~30000 
star-forming galaxies selected from the spectroscopic database of the
SDSS/BOSS DR14 \citep{Abolfathi2018} was used to construct the Mg{\sc ii}
sample.
A detailed description of the selection criteria for the extraction of galaxies with
active star formation is presented in \citet{Izotov2014}.
    Strong emission lines of H$\beta$ and
[O~{\sc iii}] $\lambda$$\lambda$4959,5007, EW(H$\beta$) $\ga$ 10 \AA\
characterise the spectra of
all selected galaxies. Galaxies with evidence of AGN features were excluded
from the sample.
   SDSS/BOSS data releases contain many more galaxies at higher redshifts
 compared with earlier data releases.
   Moreover, a wider range of the SDSS/BOSS DR14 spectra with the blue edge 
   at $\sim$3600\AA\ also allows us  to select
   galaxies with lower redshifts compared to our earlier study
\citep{Guseva2013}. The redshift of $z$ $\ga$ 0.3 permits us to detect the 
Mg {\sc ii} $\lambda$2797, $\lambda$2803 emission doublet in the spectrum.
   Thus, out of $\sim$30000
low-metallicity star-forming galaxies, more than 4100 Mg {\sc ii}
emitters were extracted, increasing the sample size by about two orders of magnitude
compared to the study by \citet{Guseva2013}.
 The redshift distribution of the Mg{\sc ii} sample corresponds to the range
of $\sim$0.3 -- 1.05 (see Fig.~\ref{fig2}).

The line fluxes and the errors of the fluxes were derived using the
IRAF\footnote{IRAF is the Image 
Reduction and Analysis Facility distributed by the National Optical Astronomy 
Observatory, which is operated by the Association of Universities for Research 
in Astronomy (AURA) under cooperative agreement with the National Science 
Foundation (NSF).} SPLOT routine.
The line fluxes were corrected simultaneously for both reddening and underlying
hydrogen stellar absorption  in the iterative procedure described by
\citet{ITL94}. 
   The reddening effect was taken into account adopting the extinction curve of
\citet{Cardelli89} in two steps: (1) line fluxes at observed wavelengths were
first corrected for the extinction in the Galaxy and (2) the internal extinction
of each of our targets was derived in the rest-frame spectrum from the Balmer
hydrogen
decrement with the average value of C(H$\beta$)=A$_V$/2.18 near 0.2--0.3, which
is typical for the   type of galaxies considered here.
The Mg {\sc ii} $\lambda$2797, $\lambda$2803 emission line fluxes were also
corrected for the underlying stellar absorption following
\citet{BruzualCharlot2003}  \citep[see details in][]{Guseva2013}.

The SDSS spectra were fitted to determine stellar
masses $M_\star$  as integrated characteristics of our sample galaxies. 
Luminosity distances for the determination of stellar masses,
H$\beta$ luminosities
and star formation rates  were obtained using a cosmological 
calculator \citep[NED,][]{Wright2006}, based on the cosmological 
parameters $H_0$=67.1 km s$^{-1}$Mpc$^{-1}$, $\Omega_\Lambda$=0.682, 
$\Omega_m$=0.318 \citep{Planck2014}.

%%%%%%%%%%%%%%%%%%%%%%%%%%%%%%%%%%%%%%%%%%%%%%%%%%%%%%%%%%%%%

\begin{figure*}
\hbox{
 \hspace*{1.5cm}\includegraphics[angle=-90,width=0.7\linewidth]{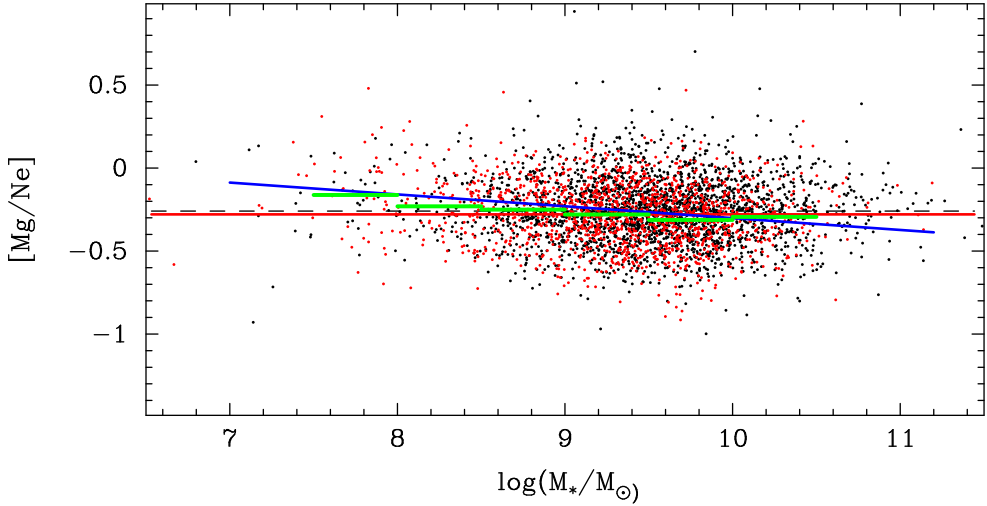}
}
\caption{[Mg/Ne] -- mass of stellar population log($M_\star$/$M_\odot$)
  relation. Symbols are as in Fig.~\ref{fig1}. A linear regression to all
  Mg {\sc ii} galaxies
  is shown by the solid blue line, and the average values in bins of stellar masses
  are represented by solid green lines. 
}
\label{fig4}
\end{figure*}

%%%%%%%%%%%%%%%%%%%%%%%%%%%%%%%%%%%%%%%%%%%%%%%%%%%%%%%%%%%%%

\section {Element abundances \label{S3}}

To determine heavy element abundances in low-metallicity galaxies by
the direct $T_{\rm e}$ method, we generally follow the prescriptions of
\citet{ITL94,ITL97,Izotov2006}.
    The direct $T_{\rm e}$ method for
    $T_{\rm e}$(O~{\sc iii}) determination is
    generally used    if measurable
[O~{\sc iii}] $\lambda$4363 emission is present. Then 
$T_{\rm e}$(O~{\sc iii}) is derived from the line ratio
[O~{\sc iii}] $\lambda$4363/$\lambda$(4959+5007).
  In the case when
[O~{\sc iii}] $\lambda$4363 is not detected we estimate 
$T_{\rm e}$(O~{\sc iii}) by the semi-empirical method by \citet{IT2007} using
strong oxygen emission lines.
For $T_{\rm e}$(O~{\sc ii}) the relation
between $T_{\rm e}$(O~{\sc iii}) and $T_{\rm e}$(O~{\sc ii}) from photoionisation
models of H~{\sc ii} regions is adopted 
\citep[see e.g.][]{Izotov2006,StasinskaIzotov2003}, 
for the direct and for the semi-empirical method.
   For ionic and total oxygen and neon abundance determinations we follow the
\citet{Izotov2006} prescription. 
\citet{Guseva2013} calculated the relation between $T_{\rm e}$(Mg~{\sc ii})
and
$T_{\rm e}$(O~{\sc ii}) using the CLOUDY photoionisation H~{\sc ii} region
models.
 The fit to this relation gives $T_{\rm e}$(Mg~{\sc ii})
 (see equation 2 in their paper), whereas
 Mg$^+$ and total Mg abundances were obtained from their equations
 3, 4, and 5.
 The heavy element abundances were derived 
using the direct $T_{\rm e}$ method in more than 35\% of our sample.
    The solar magnesium and oxygen abundances 
of \citet{A09} were adopted as a reference following the 
recommendations of \citet{Lodders2009} on whether to choose 
photospheric abundances, meteoritic abundances, or an 
average of the two values.

%%%%%%%%%%%%%%%%%%%%%%%%%%%%%%%%%%%%%%%%%%%%%%%%%%%%%%%%%%%%%

\begin{figure*}
\hbox{
  \hspace*{-0.5cm}\includegraphics[angle=-90,width=0.48\linewidth]{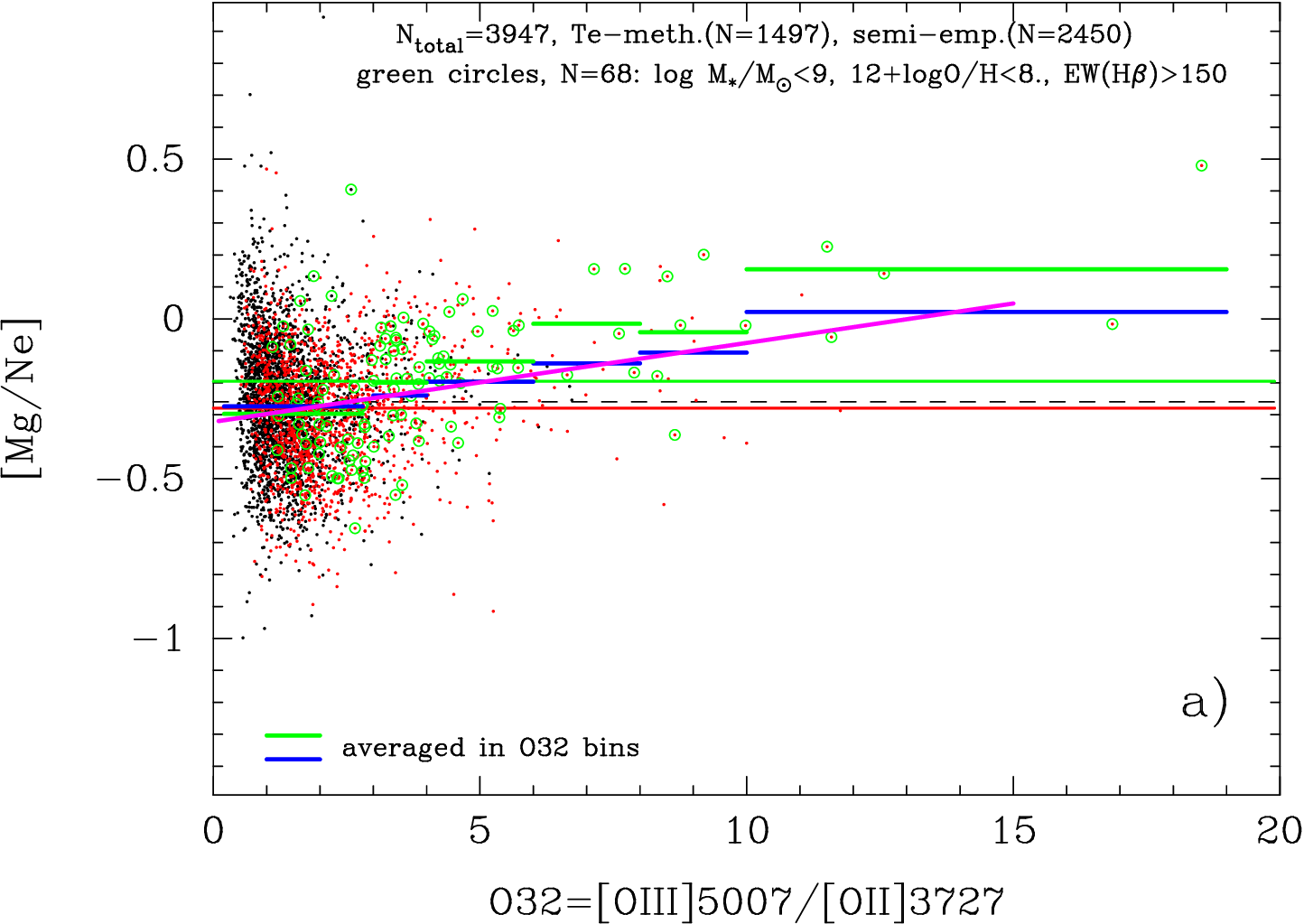}
  \hspace*{0.cm}\includegraphics[angle=-90,width=0.48\linewidth]{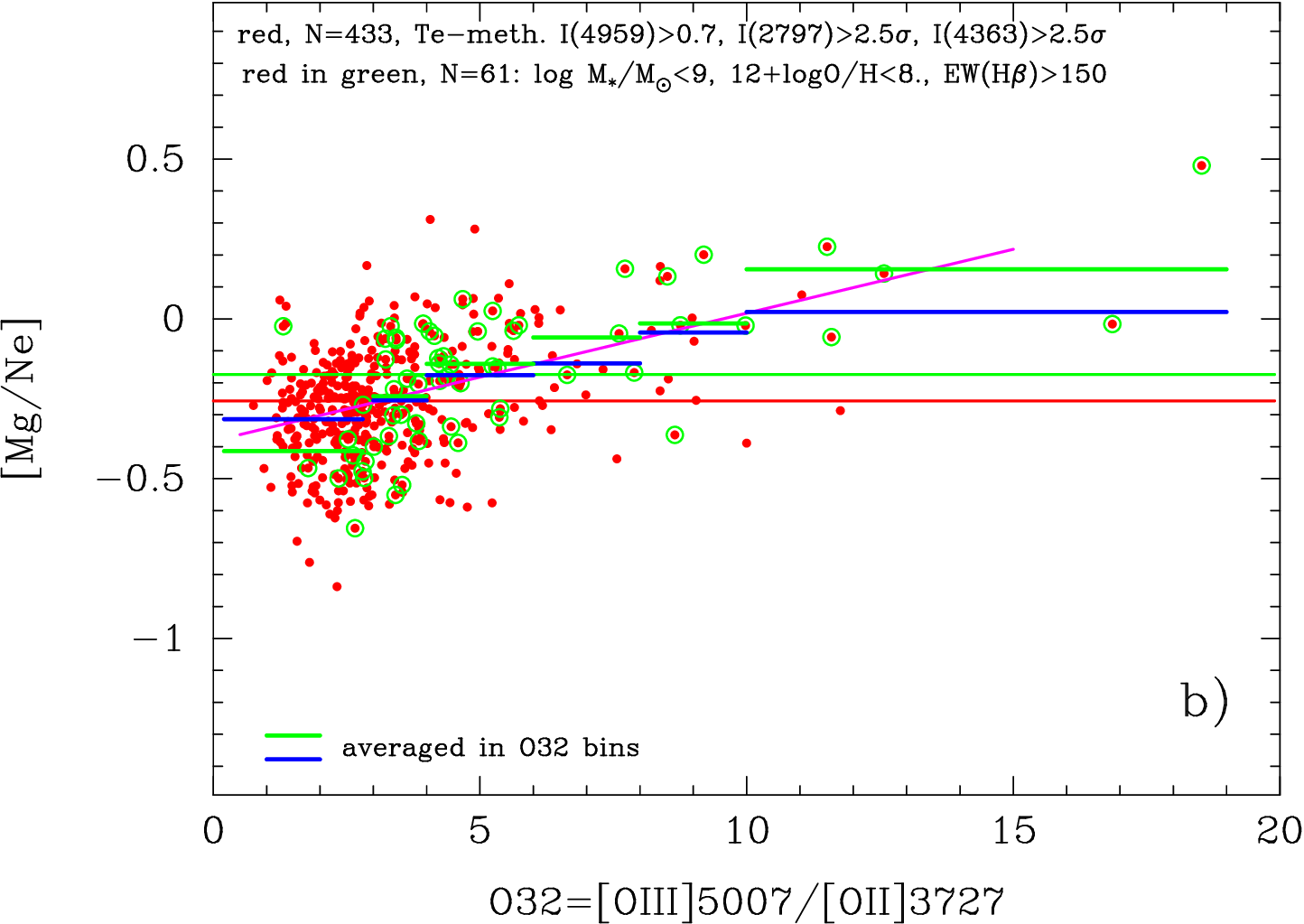}
}
\caption{\textbf{a}) [Mg/Ne] -- O$_{32}$ relations for all Mg~{\sc ii} emitters
  depicted as in Fig.~\ref{fig1}. 
  The dots inside  green circles represent low-metallicity
(12 + log O/H $<$ 8.0) low-mass (log($M_\star$/$M_\odot$) $<$ 9.0) galaxies with
high EW(H$\beta$) $>$ 150\AA.
    Linear regression to all the galaxies is shown by a purple line and average
  values in bins of O$_{32}$ are shown by blue lines for all galaxies and by
  green lines for galaxies of low metallicity, low mass, and high EW(H$\beta$). 
  \textbf{b}) The same as in \textbf{a}), but only for galaxies
  with abundance determination by the direct T$_e$ method and accuracy better
  than 40\% (more than 2.5$\sigma$) for weak but important
  Mg {\sc ii} $\lambda$2797\AA\ and [O {\sc iii}]$\lambda$4363\AA\ emission
  lines. Galaxies with flux ratios $I$[O {\sc iii}]$\lambda$4959/$I$(H$\beta$)
  less than 0.7 were also excluded, resulting in a significantly reduced number
  of galaxies N=433 (red dots).
  A linear regression to the T$_e$ data is shown by a purple line and
  average values in bins of O$_{32}$ are represented by blue lines. 
  As in \textbf{a}) the galaxies with stellar
masses log($M_\star$/$M_\odot$) $<$ 9.0, oxygen abundances 12 + logO/H $<$ 8.0,
and EW(H$\beta$) $>$ 150\AA\ are represented by dots
inside green circles.
    Average values of these galaxies in O$_{32}$ bins are shown by
    green lines. Long red and green horizontal lines are average values
    belonging to samples of corresponding colours.
}
\label{fig5}
\end{figure*}

%%%%%%%%%%%%%%%%%%%%%%%%%%%%%%%%%%%%%%%%%%%%%%%%%%%%%%%%%%%%%

\begin{figure}
\hspace*{0.cm}\includegraphics[angle=-90,width=0.96\linewidth]{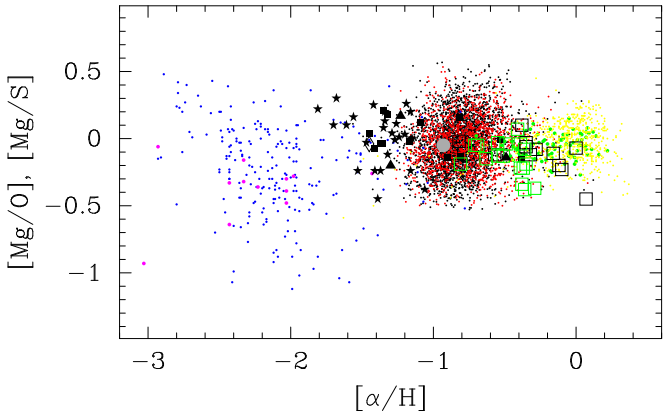}
\caption{[Mg/O] -- [$\alpha$/H] relations.
   [Mg/O] vs. [O/H] for star-forming galaxies from this paper, corrected for dust
   depletion ([Mg/O] + 0.26 for the
semi-empirical and [Mg/O] + 0.28 for the T$_e$ method)
are shown by red and black dots, as in Fig.~\ref{fig1}\textbf{a}.
All data for stars collected from the literature are presented in
[Mg/O] -- [O/H]
 relations. More specifically, for halo and disk Galactic stars from
 \citet{Roederer2014} (blue dots),
 \citet{Lai2008} (purple dots), \citet{Chen2000} (green dots), and
 \citet{Bensby2014} (yellow dots) are also presented.
Giant stars in the Milky Way satellite dwarf galaxies by
 \citet{Bonifacio2004,Sbordone2007} are shown by large black
 and green open squares, respectively. 
 Red giant branch stars in the centre of the Sculptor dwarf galaxy by
 \citet{Hill2018} are shown by black filled stars.
 [Mg/O] vs. [O/H] or [Mg/S] vs. [S/H] in the interstellar medium is
represented by DLAs
\citep[filled black squares,][]{Cia2016}, by GRB-DLAs
\citep[filled black triangles,][]{Wiseman2017},  and averaged over many DLAs \citep[filled grey circle,][]{Guseva2013}. 
}
\label{fig6}
\end{figure}

%%%%%%%%%%%%%%%%%%%%%%%%%%%%%%%%%%%%%%%%%%%%%%%%%%%%%%%%%%%%%

\section {Results \label{S4}}

Depletion onto dust of some refractory elements can be studied using
correlations
between ratios of refractory or possibly refractory  to non-refractory
elements, and metallicity or some indicator of metallicity.
The higher depletion of a refractory element is manifested in the
steepening of the 
slope of the relation between relative abundances and metallicity, for example,
in a decrease of Mg/O with increasing
metallicity, indicating that at higher metallicity a larger fraction of
magnesium is incorporated into dust.

To find the dependence of magnesium depletion 
on metallicity, we
plot a magnesium-to-oxygen abundance ratio
Mg/O relative to solar value [Mg/O] = log(Mg/O) -- log(Mg/O)$_{\odot}$
vs. 12 + log(O/H) (see Fig.~\ref{fig1}a).
   The average  value of
[Mg/O] calculated for all data regardless of the methods used for temperatures
$T_{\rm e}$(O~{\sc iii}) and $T_{\rm e}$(O~{\sc ii}), and oxygen abundance
determination is --0.26.
    We note  that for the
direct $T_{\rm e}$ method and for the semi-empirical method the average values of
[Mg/O] are very similar, --0.28 and --0.26, respectively.
    There is no appreciable
trend in  Fig.~\ref{fig1}a, which  means that
there is no dependence of magnesium depletion (relative to that of oxygen) on
metallicity, at least not
in the metallicity range of 12 + logO/H $\simeq$ 7.3 -- 8.6
(--1.4 $\la$  [O/H] $\la$ --0.1).
  This may indicate that similar fractions of Mg and O are locked in dust.

 A small fraction of oxygen is, among other candidate
 constituents, part of the interstellar dust.
     Thus, \citet{Esteban1998}, assuming a certain composition for
 the dust grains
 containing oxygen and the three most depleted elements (Mg, Si, and Fe)
 evaluated the fraction of oxygen embedded
 in dust grains.
     They proposed a correction
factor equal to 1.2 (or 0.08 dex for log O/H) for the oxygen abundance derived
from nebular spectra, taking into account the oxygen depletion onto grains.
      According to \citet{Esteban1998} the gas-phase fraction of oxygen is
 near 80\%,  at least for the Orion nebula.
     The oxygen depletion factor at solar metallicity of --0.22 is used
 in the models of \citet{Dopita2005} and in the paper of
 \citet{KewleyDopita2002}. 
    Later \citet{Dopita2013} proposed an updated version of \citet{Dopita2005} and
 adopted an oxygen depletion factor of --0.07 dex. 
 \citet{PP2010} and \citet{Mesa-Delgado2009} accepted a depletion factor
 approximately --0.1 dex.
      \citet{DiazStasinska2011} assumed a depletion factor  $\simeq$ --0.15 dex
for the Orion star-forming regions.      
     More recently \citet{Cirpiano2017}, following \citet{Bresolin2016},
 recommended to use --0.1 dex for the correction of measured gas phase
 oxygen abundances due to its depletion onto dust grains. 
    \citet{Izotov2006} derived an increase of the Ne/O abundance ratio by
$\sim$0.1 dex in the 12 + log O/H range of 7.1 and 8.6. They interpreted
this dependence suggesting zero depletion of oxygen at the lowest metallicity and
$\sim$0.1 dex depletion at solar metallicity.
   Although the precise values for the amount of oxygen locked up in dust have not been
clearly established, a fraction of 10 -- 20\% of oxygen can be incorporated
in interstellar dust grains.

%%%%%%%%%%%%%%%%%%%%%%%%%%%%%%%%%%%%%%%%%%%%%%%%%%%%%%%%%%%%%

\begin{figure*}
\hbox{
 \hspace*{-0.1cm}\includegraphics[angle=-90,width=0.5\linewidth]{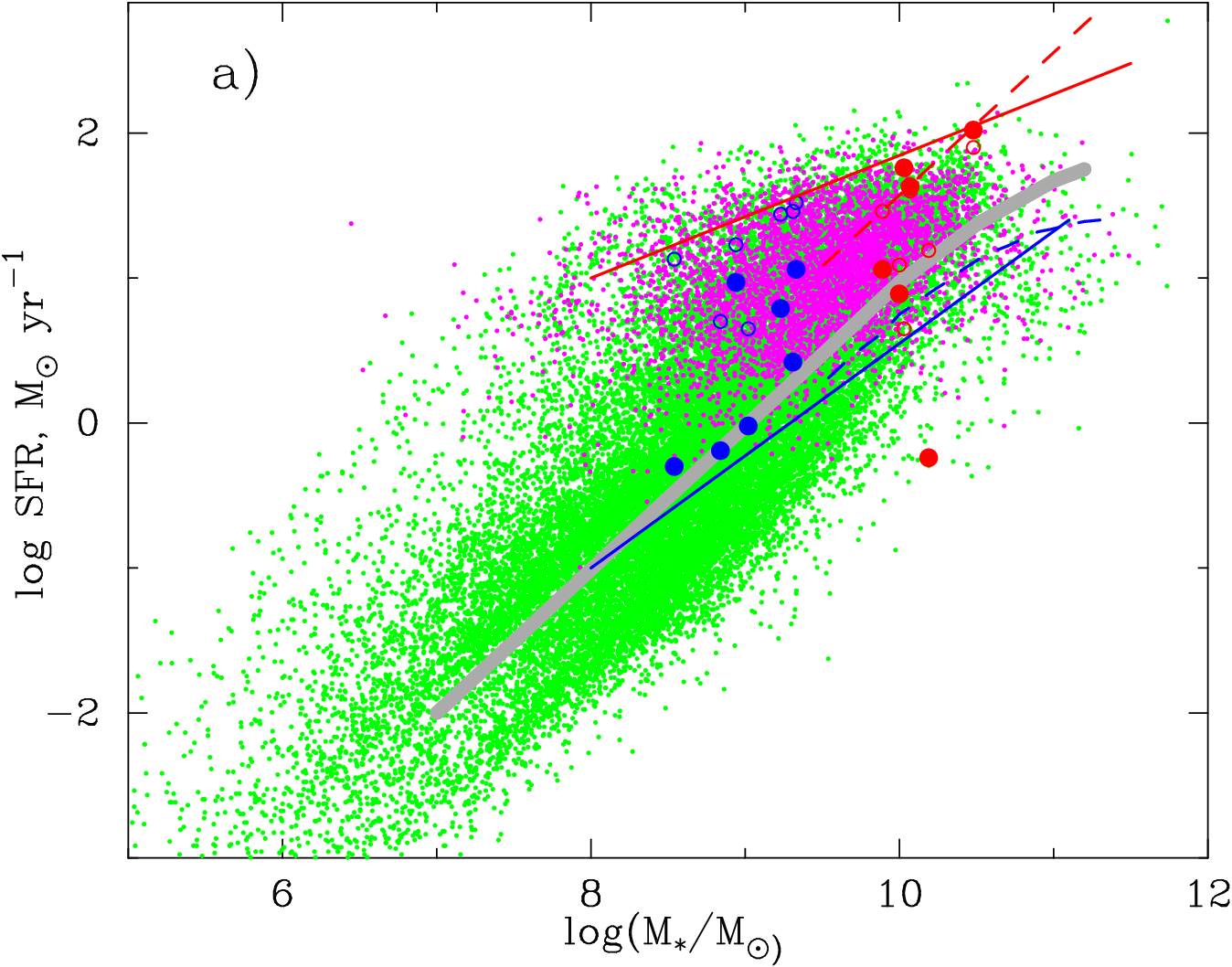}
 \hspace*{0.0cm}\includegraphics[angle=-90,width=0.5\linewidth]{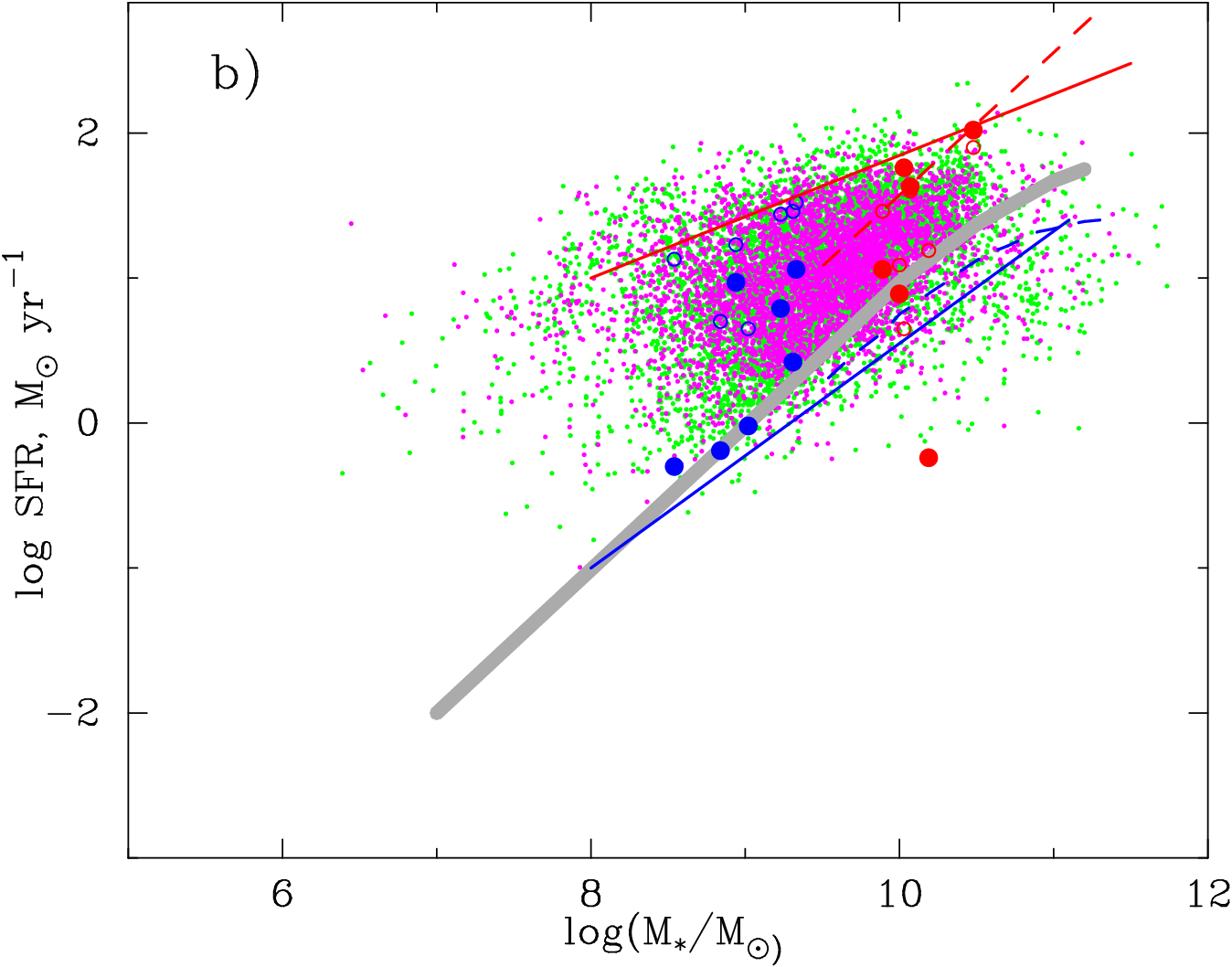}
}
\caption{\textbf{a}) log SFR vs. log($M_\star$/$M_\odot$) for our Mg~{\sc ii}
  sample (purple dots).
    Additionally, the entire SDSS DR14 sample of low-metallicity star-forming
    galaxies ($\sim$30000) is shown by green dots.
    The same is also shown in \textbf{b}),  
 but galaxies with redshift
 $z$ $<$ 0.3 are excluded from the entire SDSS DR14 sample.
    Mg {\sc ii} and Fe {\sc ii}$^*$ emitters
from the Ultra Deep Field, UDF-10, by \citet{Finley2017}  are shown by
large symbols.
     In particular, Mg {\sc ii}
  emitting galaxies are denoted
  by large blue filled circles (where SFR values are obtained from SED fitting)
  and by open circles (where SFR values are obtained from the luminosity of
  [O {\sc ii}] lines).
  Thick grey lines represent the main sequence of star-forming galaxies from
  \citet{Finley2017},  who used the results of
\citet{Schreiber2015} and \citet{Whitaker2014} and extrapolated their
data into a parameter space with no data for low $M_\star$.
    For comparison, we also show the observed data and
 extrapolations to lower masses for star-forming galaxies from
 \citet{Whitaker2012} ([0.0 $<$ $z$ $<$ 0.5] and [2.0 $<$ $z$ $<$ 2.5]).
 In both panels the solid red and blue lines represent the relations for
 high- and low-redshift galaxies by \citet{Whitaker2012}.
 We also plot the data and extrapolations from \citet{Schreiber2015}
  by dashed red [3.5 $<$ $z$ $<$ 5.0] and dashed blue [0.3 $<$ $z$ $<$ 0.7]
  lines.
} 
\label{fig7}
\end{figure*}

%%%%%%%%%%%%%%%%%%%%%%%%%%%%%%%%%%%%%%%%%%%%%%%%%%%%%%%%%%%%%

\begin{figure*}
\hbox{
 \hspace*{-0.1cm}\includegraphics[angle=-90,width=0.5\linewidth]{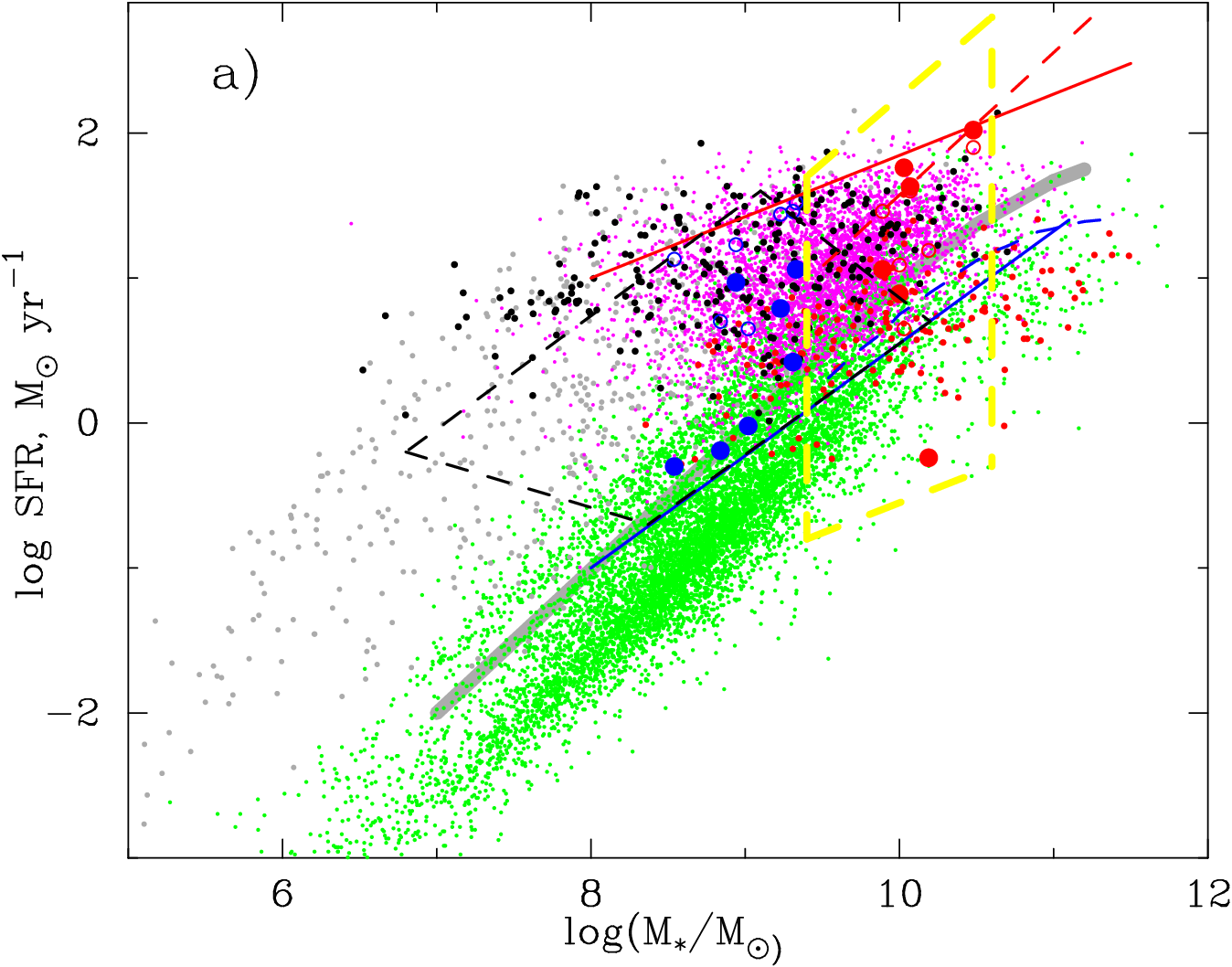}
 \hspace*{0.0cm}\includegraphics[angle=-90,width=0.5\linewidth]{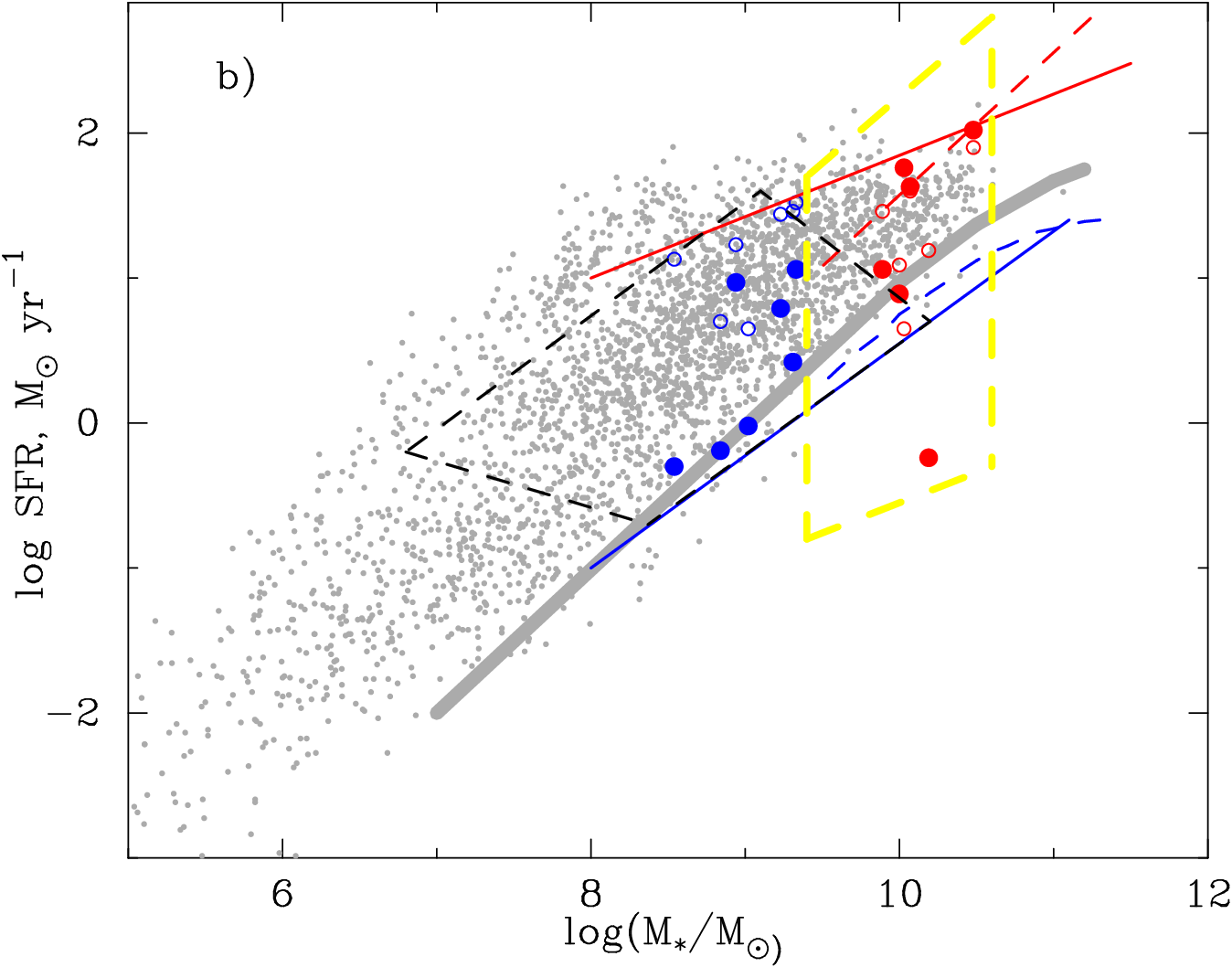}
}
\caption{\textbf{a}) log SFR vs. log($M_\star$/$M_\odot$) 
  for our entire Mg~{\sc ii} sample (purple dots);  the magnesium emitting
  galaxies with equivalent widths of H$\beta$
  emission lines EW(H$\beta$) $>$ 200\AA\ and $<$ 20\AA\ are emphasised by
  \textbf{small}
 black and red dots, respectively.
   Additionally, out of our entire SDSS DR14 sample of low-metallicity star-forming
  galaxies ($\sim$30000) only galaxies with EW(H$\beta$) $>$ 200\AA\ and 
  EW(H$\beta$) $<$ 20\AA\ are shown by grey and green dots,
  respectively.
  We also include the observed data and extrapolations to lower masses for
  star-forming galaxies from \citet{Whitaker2012}  [0.0 $<$ $z$ $<$ 0.5] by a 
  solid blue line and [2.0 $<$ $z$ $<$ 2.5] by a solid red line.
    Additionally, the 
  locations of Mg {\sc ii} emitters by \citet{Feltre2018} are indicated by
  dashed black lines and their Mg {\sc ii} absorbers by dashed yellow lines.
  Other symbols are the same as in Fig.~\ref{fig7}.
  In \textbf{b})
  only SDSS DR14 star-forming galaxies with
  EW(H$\beta$) $>$ 200\AA\ are shown. Symbols are as in \textbf{a}).
}
  \label{fig8}
\end{figure*}
%%%%%%%%%%%%%%%%%%%%%%%%%%%%%%%%%%%%%%%%%%%%%%%%%%%%%%%%%%%%%%

\begin{figure}
 \hspace*{0.0cm}\includegraphics[angle=-90,width=0.96\linewidth]{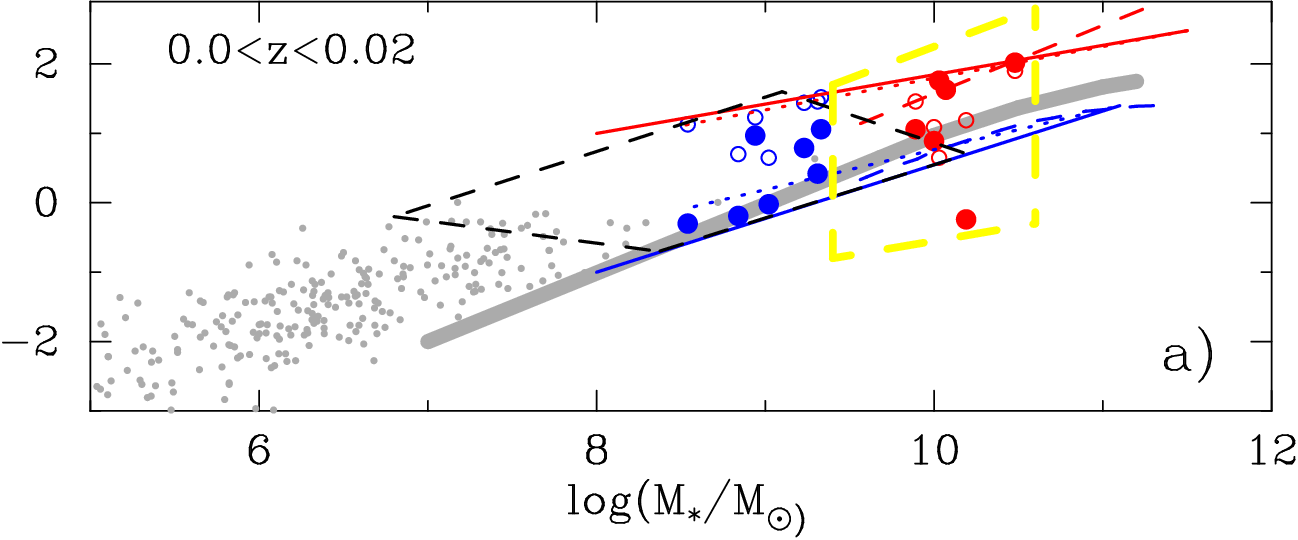}
 \hspace*{0.0cm}\includegraphics[angle=-90,width=0.96\linewidth]{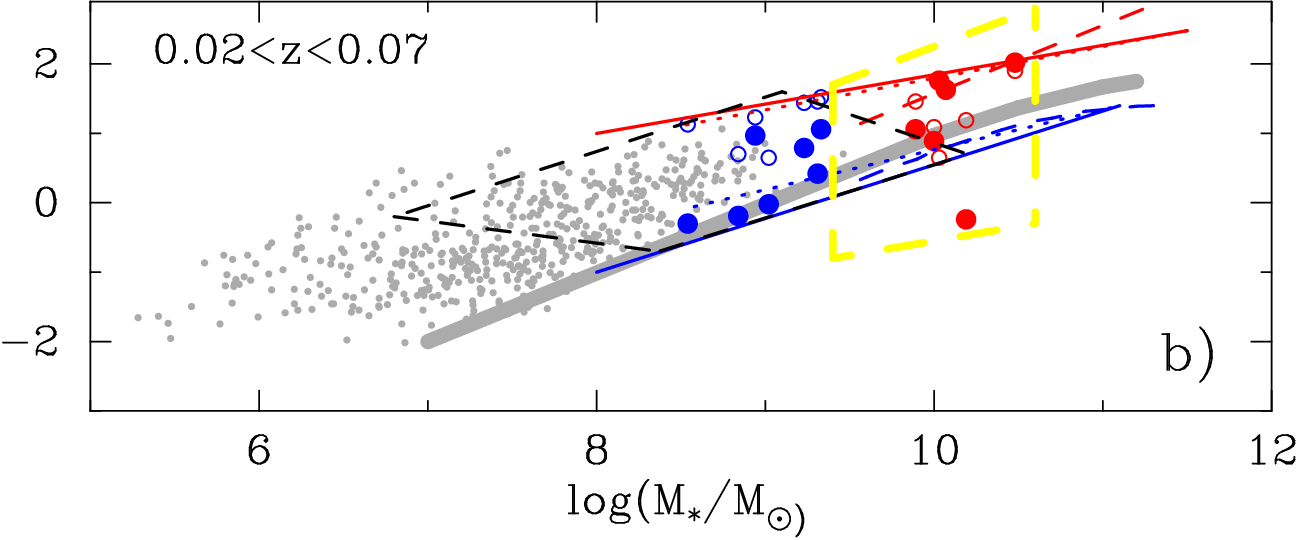}
 \hspace*{-0.63cm}\includegraphics[angle=-90,width=1.03\linewidth]{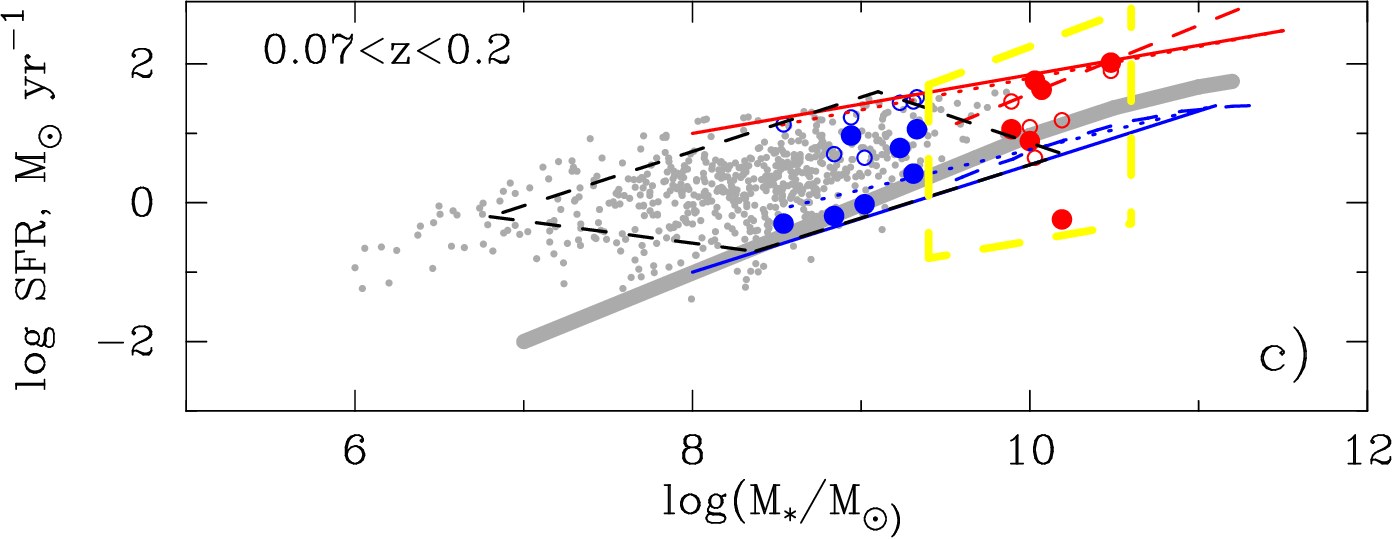}
  \hspace*{0.0cm}\includegraphics[angle=-90,width=0.96\linewidth]{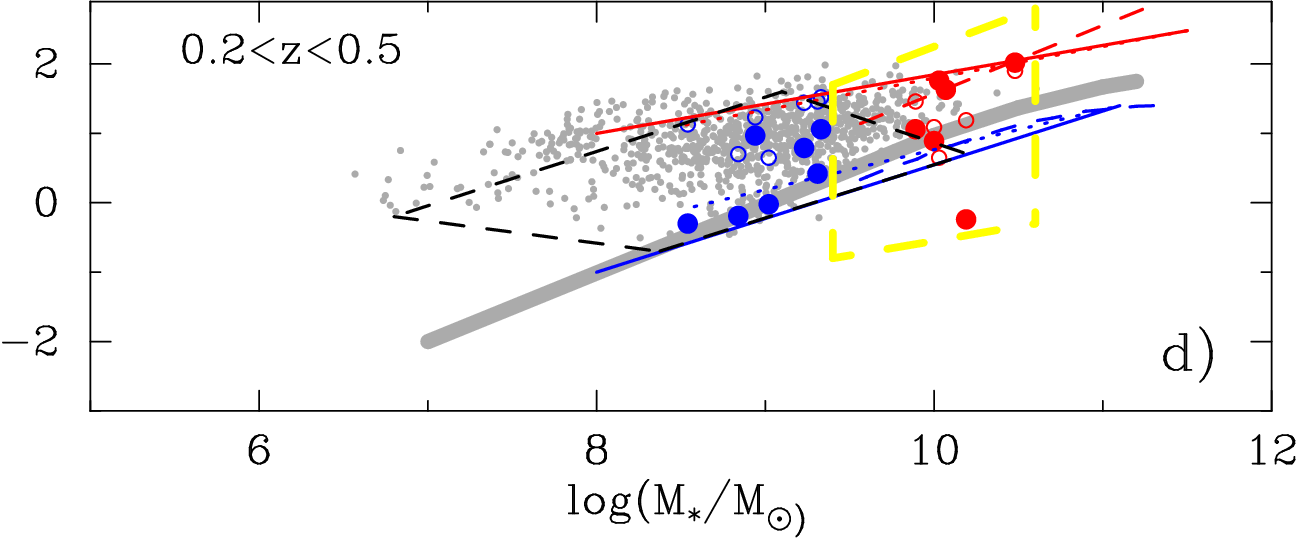}
  \hspace*{0.0cm}\includegraphics[angle=-90,width=0.96\linewidth]{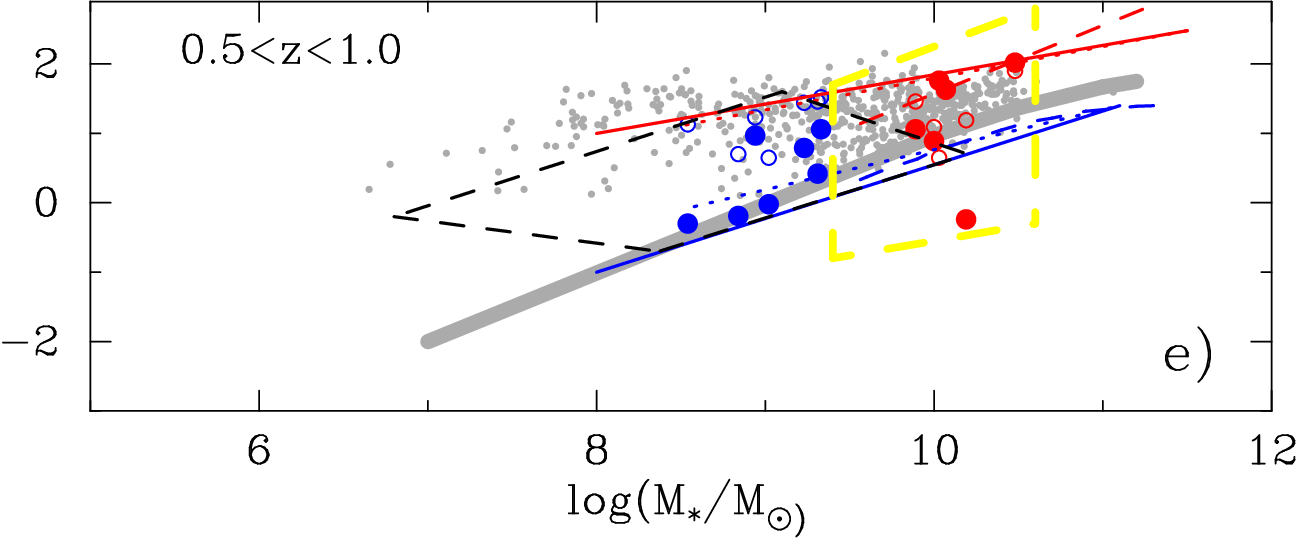}
  \caption{log SFR vs. log($M_\star$/$M_\odot$) for SDSS DR14 star-forming
    galaxies 
    with EW(H$\beta$) $>$ 100\AA\ in bins of redshifts. All
symbols and lines are as in Fig.~\ref{fig8}.
}
  \label{fig9}
\end{figure}
%%%%%%%%%%%%%%%%%%%%%%%%%%%%%%%%%%%%%%%%%%%%%%%%%%%%%%%%%%%%%

A hint supporting this is the slight slope of the dependence of the neon-to-oxygen
 ratio on metallicity in Fig.12 by \citet{Guseva2011}.
  The figure was obtained
  using   a large precise database collected from a sample for the primordial
  He abundance
 determination \citep[HeBCD sample,][]{IT2004,Izotov2004} and based on  data from
 the SDSS DR3 \citep{Izotov2006}.
    The same effect is seen in Fig.~\ref{fig1}b (this paper) for our large
 Mg {\sc ii} sample selected from SDSS DR14.
     Because neon is a noble gas this trend might be explained assuming
that some amount of oxygen 
is likely coupled to dust, which led us  to use neon instead of oxygen in our study of
magnesium dust depletion.
In agreement with the chemical evolution model predictions for Mg and Ne by
  \citet{Prantzos2018} (their Fig. 13) for both rotating and non-rotating
  massive star yields, we assume here that the intrinsic [Mg/Ne] value
  is a constant
 at least in the range of --1.5 $<$ [O/H] $<$ 0.0. 
 In Fig.~\ref{fig1} we show the correlation between [Mg/Ne] and the 
metallicity expressed either in 12 + log(O/H) (Fig.~\ref{fig1}c)  or 
in 12 + log(Ne/H) (Fig.~\ref{fig1}d). 
Some small trends are present in these plots with almost the same average
levels of depletion. The linear regression to all data (black solid line
in Fig.~\ref{fig1}d) is as follows:

 \begin{equation}
[\rm Mg/Ne] = 0.8849\pm0.0890 - 0.1606\pm0.0125(12 +\rm logNe/H).  \label{eq1}    
 \end{equation}Thus, [Mg/Ne] = 0.0 at  [Ne/H] $\simeq$ --2.4, whereas
   [Mg/Ne] = --0.388 at solar metallicity ([Ne/H] = 0.0).

 In Fig.~\ref{fig2}a the relation between [Mg/Ne] and redshift is plotted.
  We do not find any obvious trend of [Mg/Ne] with redshift.
  Metallicity is also uniformly distributed with redshift (Fig.~\ref{fig2}b).
  This means that  for our sample we do not expect evolutionary effects in
the range between $z$ $\sim$0.3 and $\sim$1.
   The total extinction A$_V$ in $V$ band derived from the extinction
coefficient C(H$\beta$) varies from 0.0 to more than 1.0, but [Mg/Ne] remains
unchanged at any A$_V$ (Fig.~\ref{fig3}).

\subsection {Dependence of [Mg/Ne] on stellar mass \label{S4-1}}

Because the [Mg/Ne] ratio exhibits a small trend with metallicity
(section \ref{S3}, Figs.~\ref{fig1}c and d) we 
may expect a similar
trend of the [Mg/Ne] ratio with galaxy stellar masses:  the stellar masses of galaxies 
correlate with metallicities through
a well-known mass--metallicity relation
\citep[see e.g.][]{Tremonti2004,Lee2006,Maiolino2008,Manucci2010,Amorin2010,Zahid2011,Zahid2012,Zahid2013,Steidel2014,Maier2014,Izotov2015,Guseva2017},
in the sense that the more massive galaxies are more metal abundant.
    The presence of such a trend is well established
even though  different types of galaxies and
adopted methods give different slopes in the relation.

Stellar masses were derived following detailed prescription by 
\citet{Guseva2006,Guseva2007} and \citet{Izotov2011}.
    Briefly, adjusting  the modelled spectral energy distributions (SEDs)
to the observed
spectra we were able to derive stellar masses after subtraction of nebular
continuum
and nebular emission lines from the observed spectra in the entire
wavelength range $\lambda$$\lambda$3600-10300\AA\ of the SDSS DR14 spectra. 
  We calculated a series of model SEDs and chose the best fit
from $\chi^2$ minimisation of the
deviation between the observed and the modelled continuum.
  Star formation history was modelled 
by a short burst of star-forming galaxies  with age less than 10 Myr, which represents
the young stellar population in addition to continuous star formation
responsible for older stars with ages $t_1$ and $t_2$ ($t_1$ $<$ $t_2$)
randomly varied in the range
between 10 Myr and 15 Gyr. The derived age $t_2$ of the oldest stars
is $<$ 10 Gyr in most of galaxies.

  The relation between [Mg/Ne] and stellar mass
is shown in Fig.~\ref{fig4}.
  The linear regression to all  Mg {\sc ii} galaxies
is shown by a solid blue line and average values in bins of stellar masses
by solid green lines. 
    A slight slope is seen, meaning that more
massive galaxies have higher magnesium depletion.
In particular, [Mg/Ne] $\simeq$ --0.15 
for low-mass galaxies with
$M_\star$ $\simeq$ 3 $\times$ 10$^7$$M_\odot$  and $\simeq$ --0.35
for high-mass galaxies with $M_\star$ $\simeq$ 3 $\times$ 10$^{10}$$M_\odot$.
A similar behaviour  is seen for the magnesium depletion, which
increases with metallicity (Fig.~\ref{fig1}c and d).

\subsection {Dependence of [Mg/Ne] on O$_{32}$ \label{S4-2}}

O$_{32}$=[O {\sc iii}]$\lambda$5007/[O {\sc ii}]$\lambda$3727 is an important
parameter to search
for galaxies with Lyman continuum (LyC) escaping radiation.
   The parameter O$_{32}$ is usually used for the selection of galaxy
candidates able to ionise the intergalactic medium,
which is interesting in connection with the problem of the reionisation
of the Universe at redshifts $z$ $\sim$ 5--10. 
  The dependence of [Mg/Ne] on O$_{32}$ is shown in Fig.~\ref{fig5}.
There is very large scatter of [Mg/Ne] for Mg {\sc ii} galaxies
with O$_{32}$ $\sim$ 0 -- 3 (Fig.~\ref{fig5}a).
Nevertheless, in galaxies with O$_{32}$ below 3 -- 4 [Mg/Ne] is constant, but at
higher O$_{32}$ [Mg/Ne]  increases with increasing O$_{32}$. 
   Linear regression to all the galaxies (purple line) and average values
in bins of O$_{32}$ (blue lines) demonstrate 
an increase in [Mg/Ne] with increasing O$_{32}$, which depends on the
ionisation parameter.
    Moreover, the galaxies with O$_{32}$ $>$ 5 are above the average value
of [Mg/Ne] (long solid red and dashed black lines).
    We note that almost all 
galaxies with O$_{32}$ $>$ 5 have low masses (log($M_\star$/$M_\odot$) $<$ 9.0),
low metallicities (12 + logO/H $<$ 8) and high EW(H$\beta$) $>$ 150\AA\
(dots inside  green
  circles). 
     In Fig.~\ref{fig5}b we only show  galaxies
 with abundances derived by the direct $T_{\rm e}$ method and accuracies better
 than 40\% (more than 2.5$\sigma$) for the weak but important
  Mg {\sc ii} $\lambda$2797\AA\ and [O {\sc iii}]$\lambda$4363\AA\ emission
  lines. Galaxies with flux ratios $I$[O {\sc iii}]$\lambda$4959/$I$(H$\beta$)
  less than 0.7 were also excluded to ensure that we use only
 high-excitation galaxies whose [O {\sc iii}]$\lambda$4363\AA\
 emission lines are reliably detected. The trend of increasing [Mg/Ne] with
  increasing O$_{32}$ becomes even more evident (purple line in
  Fig.~\ref{fig5}b).
    
  Such a strong trend in Fig.~\ref{fig5} cannot be explained by metallicity
effects because there is no dependence on metallicity in the models of
\citet{Prantzos2018} (see also Fig.~\ref{fig6} in this paper),
especially in the
range of [O/H] from 0.0 to --1.5 dex occupied by our Mg {\sc ii} sample.
     It is also found in the present paper that the dependence of [Mg/Ne] on
metallicity is not too strong (see Fig.~\ref{fig1}d).
   Let us consider possible reasons for this effect:
1) density bounded models are often adopted to explain the large O$_{32}$
in star-forming galaxies \citep{JaskotOye2013,Nakajima2016,Izotov2017}. We can also take  into
account that the Mg ionisation potential of 7.65 eV is much lower than that for
Ne (21.56 eV), 
   but if this is the case 
[Mg/Ne] must be even lower for high O$_{32}$ galaxies; 
2) efficient warming-up and possible destruction of interstellar dust grains
by the intense ionising UV radiation of young stars,
which is characterised by high EW(H$\beta$)
\citep{Izotov2011a,Izotov2014,Izotov2014a}, because the number of massive O
stars reaches  10$^2$--10$^4$ in super star clusters \citep{Schaerer2000}.
   Then Mg as a refractory element could not be locked in the dust and
depletion of the magnesium would be near zero;
and    3) uncertainties in the ionisation correction factor $ICF$(Mg$^+$) at high
O$_{32}$ corresponding to O$^{2+}$/O $>$ 0.9
\citep[see for details ][]{Guseva2013}.

\subsection {Comparison with data from the literature\label{S4-3}}

\textit{[Mg/O] vs. metallicity [$\alpha$/H]}.
In Fig.~\ref{fig6} the relation between
[Mg/O]  and
[O/H] is given for our data and other data collected from the
literature in a wide range of metallicity.
Only $\alpha$ elements (Mg, O, and S) are used for the sake of a direct
comparison of different data.
   \citet{Roederer2014} (blue dots) obtained element abundances
for more than 300 very metal-poor Galactic halo stars down to
[O/H] $\sim$ --3.0 dex.
    We selected 240 stars
with precise measurements of magnesium and oxygen abundances.
Out of 28 very low-metallicity stars from \citet{Lai2008}, 15 stars
in a metallicity range [O/H] from
--2 to --3 dex are shown in purple and
   78 Galactic disk stars out of 90 from \citet{Chen2000}
with --0.7 $<$ [O/H]  $<$ +0.2 dex are plotted in green.
   \citet{Bensby2014} studied a large sample of 714 Galactic disk and thin
   disk stars with nearly solar [O/H] (yellow
   dots).
Giant stars in the Milky Way satellite dwarf galaxies by
\citet{Bonifacio2004,Sbordone2007}
are shown by large black
 and green open squares, respectively. 
 Red giant branch stars in the centre of the Sculptor dwarf galaxy by
 \citet{Hill2018} are shown by black filled stars.
 
Abundances of Mg, S, and O  are given for DLAs in
the paper of \citet{Cia2016}, \citet{Wiseman2017}, and averaged over
  many DLAs in \citet{Guseva2013}.
These data
represent the relative
abundances obtained for the interstellar medium (filled black squares and
triangles, and large filled grey circle,
respectively). To correct our
Mg sample for dust depletion, the [Mg/O] for each galaxy was shifted by +0.28
or +0.26 depending on whether the $T_{\rm e}$ or the semi-empirical method
was used (Fig.~\ref{fig1}a).
  The [Mg/O] values reported here are in good agreement with similar
measurements made in stars and DLAs. There is no appreciable trend of the [Mg/O]
ratio with metallicity  for all of the data presented in Fig.~\ref{fig6}.

\textit{SFR vs. M$_\star$}.
The star formation rate  is one of the regularly used parameters of
star-forming activity of a galaxy, and  is derived from the H$\alpha$ luminosity
$L$(H$\alpha$) following \citet{Kennicutt1998}.
   The behaviour of star-forming galaxies in the SFR versus stellar mass
diagram (the so-called main sequence) in terms of evolution of the
relation with increasing 
redshift has been actively investigated in recent years by many authors
\citep[see e.g.][]{Bouche2010,Karim2011,Whitaker2012,Whitaker2014,Schreiber2015,Mitra2017}.
One of the conclusions from these studies is that the main sequence is almost
linear and does not demonstrate a significant evolution with redshift from
lowest masses up to
$M_\star$ $\sim$ 10$^{9.5}$ $M_\odot$.
   In Fig.~\ref{fig7} we 
   show the position of our Mg {\sc ii} sample (purple dots) and entire
   SDSS DR14 sample of low-metallicity
star-forming galaxies ($\sim$30000, green dots)
on the log SFR -- log($M_\star$/$M_\odot$) diagram
and compare it with the corresponding distributions
obtained by others authors.

About three hundred Mg {\sc ii} and Fe {\sc ii}$^*$ emitting galaxies,
which were identified by \citet{Finley2017} with the VLT/MUSE in the
$Hubble$ Ultra Deep Field South \citep{Bacon2015},
are located along the galaxy main sequence, but their
distribution shows a dichotomy such that the galaxies with masses
log($M_\star$/$M_\odot$) $<$ 9.0 are Mg {\sc ii} emitters, whereas galaxies
with masses higher than 10$^{10}$ $M_\odot$  exhibit solely
Fe {\sc ii}$^*$ emission without accompanying
Mg {\sc ii} emission. 
Mg {\sc ii} and Fe {\sc ii}$^*$ emitters from the \citet{Finley2017} UDF-10
field are shown in Fig.~\ref{fig7}.
   In particular,  Mg {\sc ii} emitting galaxies
are denoted by blue filled circles (SFR values  obtained from SED
fitting) and by blue open circles (SFR values  obtained from the
luminosity of the [O {\sc ii}] lines).
  Fe {\sc ii}$^*$ emitters are represented by
the same symbols, but  in red.
    In general our entire SDSS DR14 star-forming sample, which also includes  our
Mg {\sc ii} sample (Fig.~\ref{fig7}a), is stretched along the
galaxy star formation main sequence 
from \citet{Finley2017}, but the bulk of our Mg {\sc ii}
emitters at a fixed mass is shifted to
higher SFRs (higher H$\alpha$ luminosities) or at a fixed SFR to lower masses
compared to the main sequence. 
    Only galaxies from our entire SDSS DR14 sample with redshift $z$ $>$ 0.3
are presented in Fig.~\ref{fig7}b.
  The redshift restriction \textbf{is used} because of the SDSS DR14 blue wavelength limit of
$\sim$3600\AA.
  In the absence of this limit we  could likely detect Mg {\sc ii} emitting
galaxies located above the main sequence and extended to lower masses. 
  Both Mg {\sc ii} and Fe {\sc ii}$^*$ emitting samples from \citet{Finley2017}
are located within our magnesium sample.

In Fig.~\ref{fig8} there is a clear shift between our Mg {\sc ii}
  emitters with high EW(H$\beta$) (small black dots) and those with low
  EW(H$\beta$) (small red dots).
  A similar shift between galaxies with high and low EW(H$\beta$) is seen
for the 
SDSS DR14 sample (grey and green dots).
   The latter galaxies extend to lower SFRs and lower stellar masses.
   The magnesium resonant doublet in emission and absorption was investigated
by \citet{Feltre2018} using data from the MUSE $Hubble$  Ultra Deep Field
Survey \citep{Bacon2017}.
     Exploiting photoionisation models by \citet{Gutkin2016} and their
own sample of almost 400 star-forming galaxies [0.7$<$ $z$ $\la$ 2.3], 
\citet{Feltre2018} 
found that Mg {\sc ii} emission has a nebular origin.
This conclusion verifies our earlier finding \citep{Guseva2013} obtained using
the CLOUDY code by \citet{Ferland1998}.
\citet{Feltre2018} also derived   that Mg {\sc ii} emitters have lower stellar
masses than  Mg {\sc ii} absorbers.

The Mg {\sc ii} emitting galaxies in both studies
\citep[large blue filled and open circles,][]{Finley2017} and
\citep[quadrilateral delineated by black dashed lines,][]{Feltre2018}
are located in the region preferably occupied by our Mg {\sc ii} emitters with
high EW(H$\beta$)
(black dots), whereas Mg {\sc ii} absorbers of \citet{Feltre2018}
(quadrilateral delineated by thick yellow dashed lines) as well as
Fe {\sc ii}$^*$ emitting galaxies of \citet{Finley2017} (large red
filled and open circles) 
are located in the region where our Mg {\sc ii} emitters with low EW(H$\beta$)
are preferably located (red dots).

We note  that our entire Mg {\sc ii} sample 
shows a spread in 
stellar masses from 10$^{6.5}$ to 10$^{11.5}$ $M_\odot$.
   The higher redshift [0.70 $<$ $z$ $<$ 2.34] Mg {\sc ii} emitters of
\citet{Feltre2018} have, on average, lower
masses and lower SFRs compared to our Mg {\sc ii} sample and to the Mg {\sc ii}
emitters from
\citet{Finley2017}, and
extend to low masses of $\sim$10$^{7}$ $M_\odot$ and to low SFRs of
0.2--0.3$M_\odot$ yr$^{-1}$.
   It is interesting that they are mainly located in the region of high
EW(H$\beta$) (black and grey dots).
     The emission of nebular magnesium is quite expected
     in galaxies with young age of a recent strong burst of star
formation activity, which have the lowest masses.
    Once the magnesium in the star-forming galaxies has a
nebular origin, the Mg {\sc ii} emitters could be observed in the galaxies
with high
EW(H$\beta$), i.e. at an early stage of the bursty star formation.
In principle,
Mg {\sc ii} emitters are expected even at the lowest masses and
lowest SFRs (region along the sequence of grey dots). 
   However, they are too faint and have  redshifts that are too low to be detected.
   Mg {\sc ii} emitters from \citet{Feltre2018} with higher redshift have the
same shift from the star-forming main sequence and the same slope 
as star-forming galaxies with EW(H$\beta$) $>$ 100\AA\ from this paper. 

Effects of the observational selection 
result in a smaller slope of the
galaxy star formation main sequence at the high-mass end because at higher
redshifts low-mass galaxies become too faint.
    The observed data and extrapolations to lower masses for star-forming
galaxies
from \citet{Whitaker2012} for their two limiting cases of $z$=0.0 (solid blue
line) and $z$=2.5 (solid red line) match quite well our high EW(H$\beta$)
sequence (grey and black dots in Fig.~\ref{fig8}a).
    In Fig.~\ref{fig9} the sequence of SFR vs. $M_{\star}$  with 
progressively increasing redshifts for SDSS DR14 star-forming galaxies with
EW(H$\beta$) $>$ 100\AA\ is shown.
In any redshift bin 
excluding the lowest redshift bin there is a deficiency of
galaxies with lowest masses.   
     Moreover,  this effect is greater at greater redshifts.
     The left side envelope  in each bin shows the same shift from the main
     sequence.
   Thus, we can conclude that the star-forming galaxy main sequence is universal with
the same slope at any redshift \citep[see Fig.10 in ][]{Izotov2015} if
the effect of observational selection is correctly taken into account.

\section{Summary \label{S5}}

We present the determination of the
interstellar magnesium abundance as derived from the resonance 
emission-line doublet Mg~{\sc ii} $\lambda$2797, $\lambda$2803 in 
4189 SDSS spectra of low-metallicity emission-line star-forming
galaxies with redshifts $z$ $\sim$ 0.3 -- 1.05.
 This 
emission is detected in $\sim$35\% of the entire sample of low-metallicity
star-forming galaxies with redshifts $z$ $\ge$ 0.3 selected from SDSS DR14
\citep{Abolfathi2018}.

 We study the dependence of the magnesium-to-oxygen and
magnesium-to-neon abundance ratios on metallicity.
 We use the magnesium-to-neon ratio relative to the solar
value [Mg/Ne] instead of [Mg/O] in evaluation of magnesium depletion in the
interstellar medium  because  neon is a noble gas and  does not
incorporate into dust. The dependence of [O/Ne] on metallicity is explained by
the coupling of a small amount of oxygen into dust grains. We derive
magnesium depletion of [Mg/Ne] $\simeq$ --0.4 at solar metallicity.

The global parameters of the magnesium sample such as the mass of the
stellar population,
star formation rate, and extinction coefficient C(H$\beta$) are
derived and compared with investigations of other authors.  
   More massive and more metal abundant
galaxies are found to have higher magnesium depletion.
   Our data for interstellar magnesium-to-oxygen abundance ratios
   relative to the solar value are in good agreement with
similar measurements made for Galactic
   stars,
for giant stars in the Milky Way satellite dwarf galaxies, and with
low-metallicity DLAs.

   \citet{Finley2017,Feltre2018} reported that the galaxies with Mg {\sc ii}
both in emission or in absorption are located along the star-forming galaxy main sequence
but their distribution shows a dichotomy with the dependence of SFR on
stellar mass of the galaxies.
    We show that the Mg {\sc ii} emitting galaxies from
\citet{Finley2017} and \citet{Feltre2018}
in the SFR -- $M_\star$ relation are located in the region occupied by our
Mg {\sc ii} emitters with high EW(H$\beta$),
whereas Mg {\sc ii} absorbers of \citet{Feltre2018} and Fe {\sc ii}$^*$
emitters from \citet{Finley2017} are located in the region where our Mg emitters
with low EW(H$\beta$) are preferably located. This also confirms that
Mg {\sc ii}
emission has a nebular origin. In this case the presence of emission or
absorption is determined mainly by the mass of the old stellar population and
by the age of the present burst of star formation.

\acknowledgements

N.G.G. and Y.I.I. thank  the Max-Planck 
Institute for Radioastronomy, Bonn, Germany, for the hospitality.   
They acknowledge support from the National Academy of Sciences of Ukraine
(Project No. 0116U003191). 
This research has made use of the NASA/IPAC Extragalactic Database
(NED), which is operated by the Jet Propulsion Laboratory, California
Institute of Technology, under contract with the National Aeronautics 
and Space Administration.
    Funding for the Sloan Digital Sky Survey (SDSS) has been 
provided by the Alfred P. Sloan Foundation, the Participating Institutions, 
the National Science Foundation, the U.S. Department of Energy, the National 
Aeronautics and Space Administration, the Japanese Monbukagakusho,  the 
Max Planck Society, and the Higher Education Funding Council for England.

\end{document}